\documentclass[final,5p,times,twocolumn,number]{elsarticle}
\biboptions{sort&compress}
\usepackage{soul}
\usepackage{amssymb}
\usepackage{lipsum}
\usepackage{amsmath}
\usepackage{graphicx}
\usepackage[colorlinks,linkcolor=blue,anchorcolor=blue,citecolor=blue]{hyperref}

\journal{Physics Letters B}

\begin{document}

\begin{frontmatter}
\title{Cosmic anisotropic hair of nonlocal RT gravity}
\author[aff01]{Jiajun Zhou}
\author[aff01]{Shuxun Tian}
\ead{tshuxun@bnu.edu.cn}
\author[aff01,aff02]{Zong-Hong Zhu}
\ead{zhuzh@bnu.edu.cn}
\affiliation[aff01]{organization={School of Physics and Astronomy, Beijing Normal University},
            addressline={}, 
            city={Beijing},
            postcode={100875}, 
            state={},
            country={China}}
\affiliation[aff02]{organization={School of Physics and Technology, Wuhan University},
            addressline={}, 
            city={Wuhan},
            postcode={430072}, 
            state={},
            country={China}}
\begin{abstract}
  Nonlocal RT gravity has proven effective in explaining the late-time cosmic acceleration while remaining consistent with local gravity tests. However, most previous cosmological studies of this theory have assumed an isotropic background, which may not fully capture the slight anisotropies suggested by current observations, such as those inferred from Type Ia supernovae data. In this paper, we investigate the dynamical evolution of an anisotropic Bianchi type I universe within the framework of nonlocal RT gravity. By introducing six dimensionless variables, we construct the corresponding dynamical system and perform a detailed phase-space analysis. An unexpected finding is that, contrary to many dark energy models and modified gravity theories in which anisotropies decay with time, nonlocal RT gravity predicts a growth of cosmic anisotropy. This behavior poses a challenge to the cosmic no-hair theorem within the nonlocal RT gravity scenario.
\end{abstract}
\end{frontmatter}

\section{INTRODUCTION}\label{sec:01}
The assumption of isotropy and homogeneity on large scales underpins the standard model of modern cosmology. However, recent observations, including quasars \citep{2021ApJ...908L..51S} and Type Ia supernovae \citep{2019MNRAS.486.5679Z,2019EPJC...79..783S,2023PhRvD.108l3533M}, suggest subtle hints of Bianchi anisotropy in the homogeneous cosmic background. How can the origin of such anisotropy be explained theoretically? The cosmic no-hair theorem states that a cosmological constant can wash away Bianchi anisotropy in an expanding Universe \citep{PhysRevD.28.2118}. This argument applies to both early-time inflation and late-time accelerated expansion. Most alternatives to the cosmological constant maintain this property, e.g., an exponential scalar field \citep{1992PhRvD..45.1416K,Kitada1993.CQG.10.703}. Nevertheless, there are exceptions: the inflation field can source a vector field and lead to a growing Bianchi anisotropy \citep{Watanabe2009.PRL.102.191302}. However, such a mechanism cannot account for the observed late-time Bianchi anisotropy, because the cosmic microwave background is nearly isotropic, and the radiation-dominated and matter-dominated eras within general relativity would further suppress the Bianchi anisotropy (see the flat case in \citep{1973ApJ...180..317C}). It would therefore be useful to identify a theory in which Bianchi anisotropy can be enhanced in the late-time Universe. This paper aims to analyze the dynamics of Bianchi anisotropy in nonlocal RT gravity \citep{PhysRevD.88.044033,PhysRevD.89.043008}.

As one of the prominent modifications of general relativity, nonlocal RT gravity not only elucidates the acceleration of the late-time universe \citep{Dirian_2014,PhysRevD.89.043008,Dirian_2016}, but also survives Solar System tests \citep{2014JHEP...08..029K,PhysRevD.100.124059}. The gravitational field equations are given by
\begin{equation}
G_{\mu\nu}-\frac{m^{2}}{3}\left(g_{\mu\nu}\square^{-1}_{g}R\right)^{\mathrm{T}}=8\pi G T_{\mu\nu},\label{s1}
\end{equation}
where the mass scale $m=\alpha H_0/c$  possesses the dimension of inverse length, the parameter $\alpha$ can be fixed with current observational data, and $H_0$ represents the present Hubble parameter. The inverse d'Alembertian operator $\square^{-1}_{g}$ is defined through the retarded Green function, and the superscript $\mathrm{T}$ gives the transverse part. To facilitate calculations, one generally introduces auxiliary fields $S_\mu$ and $U$ to convert the integral operator into a differential operator (see Sec. \ref{sec:02}). Previous works \citep{Dirian_2014,PhysRevD.89.043008,Dirian_2016} on cosmological fitting generally adopted FLRW background and $S_\mu=(c^2 \mathcal{S}_0, 0,0,0)$. However, the choice of  background deserves further discussion. Our previous work has discussed the cosmic background evolution and the scalar and tensor perturbations of nonlocal RT gravity with FLRW background and $S_\mu=(c^2 \mathcal{S}_0, a \mathcal{S}_1, a \mathcal{S}_1, a \mathcal{S}_1)$ \citep{PhysRevD.100.124059}. This choice of background is mathematically self-consistent. But, as stated in \citep{Belgacem_2020}, such a vector has a preferred direction in space and does not satisfy the isotropy assumption in physics. Inspired by this statement, it is necessary to generalize the FLRW metric to an anisotropic metric, which is the second motivation of this paper.
 
This work is organized as follows.
In Sec.~\ref{sec:02}, six dimensionless variables are introduced to facilitate subsequent analysis, and their dynamical equations are obtained. Sec.~\ref{sec:03} focuses on phase-space analysis. A numerical evolution result for verification will be given in Sec.~\ref{sec:04}. Conclusions are presented in Sec.~\ref{sec:06}.

\section{DYNAMIC ANALYSIS}\label{sec:02}

With the introduction of the auxiliary fields $U$ and $S_{\mu}$, Eq. (\ref{s1}) can be recast as  \citep{PhysRevD.89.043008,PhysRevD.100.124059}
\begin{gather}
    G_{\mu \nu}+\frac{m^{2}}{6}\left(2 U g_{\mu \nu}+\nabla_{\mu} S_{\nu}+\nabla_{\nu} S_{\mu}\right)=8\pi G T_{\mu \nu}\label{sc1}, 
    \\
    \square U=-R
    \label{sc2}, 
    \\
    \square S_{\mu}+\nabla^{\nu} \nabla_{\mu} S_{\nu}=-2 \partial_{\mu} U
    \label{sc3}.
\end{gather}
As discussed at the end of Sec.~\ref{sec:01}, we adopt the Bianchi type-I metric \citep{Bianchi1898}
\begin{equation}
    \mathrm{d}s^2=-c^{2}\mathrm{d}t^{2}+a^{2}_{1}\mathrm{d}x^{2}+a^{2}_{2}\mathrm{d}y^{2}+a^{2}_{3}\mathrm{d}z^{2},
\end{equation}
and for the auxiliary fields, we assume
\begin{align}
    U &=U(t), 
    \\
    S_{\mu}(t) &=\left(c^{2} \mathcal{S}_{0}, a_{1} \mathcal{S}_{1}, a_{2} \mathcal{S}_{2}, a_{3} \mathcal{S}_{3}\right)\label{ss},
\end{align}
where $\mathcal{S}_{i}$ has the dimension of length and $\mathcal{S}_{0}$ has the dimension of time.
With the above ansatz, the temporal-spatial components of Eq.~\eqref{sc1} become
\begin{equation}
    -\frac{m^{2}H_{i}}{a_{i}}\mathcal{S}_{i}+\frac{m^{2}}{a_{i}}\dot{\mathcal{S}}_{i}=0,
\end{equation}
with the solution
\begin{equation}
    \mathcal{S}_{i}=l_{i}\frac{a_{i}}{a_{i0}},
\end{equation}
where $l_{i}$ is the integration constant with the dimension of length and $a_{i0}$ is the value of $a_{i}$ at the present time. While the ratio $l_{i}/a_{i0}$ is one parameter for each $i$, we reserve both $l_{i}$ and $a_{i0}$ to facilitate the dimensional analysis in subsequent calculations.
To simplify the calculation and analysis, we introduce quantities $a,\alpha_{i},\eta,\sigma$ and $\Sigma$,
\begin{align}
    a_{i}(t) &\equiv a(t) \cdot e^{\alpha_{i}(t)},
    \label{sig1}
    \\
    (\eta, \sigma)&=\left(\alpha_{1}+\alpha_{2}, \alpha_{1}-\alpha_{2}\right),
    \label{sig2}
    \\
    \Sigma&=\left(3 \dot{\eta}^{2}+\dot{\sigma}^{2}\right) / 4,
    \label{sig3}
\end{align}
where $\alpha_{1}+\alpha_{2}+\alpha_{3}=0$ to ensure the conservation of variable numbers. From the above, it can be seen that in the following text, $\Sigma$ will reflect the anisotropy of the system. Substituting the above solution of $\mathcal{S}_{i}$ and dimensionless variables into the field equations, Eqs. (\ref{sc1}) - (\ref{sc3}) gives
\begin{subequations}
	\label{s2}
	\begin{gather} 
		3H^{2}-\Sigma-\frac{m^{2}c^{2}}{3}(U-\dot{\mathcal{S}}_{0})=8\pi G\rho,
		\\
		\frac{m^{2}c^{2}}{3}(U-H\mathcal{S}_{0})-(2\dot{H}+3H^{2}+\Sigma)=8\pi G\frac{p}{c^{2}},
		\\
		\ddot{U}+3H\dot{U}=6\dot{H}+12H^{2}+2\Sigma,
		\\
		\ddot{\mathcal{S}}_{0}+3H\dot{\mathcal{S}}_{0}-(3H^{2}+2\Sigma)\mathcal{S}_{0}=\dot{U},
		\\
		\ddot{\sigma}+(3H-\frac{m^{2}c^{2}}{3}\mathcal{S}_{0})\dot{\sigma}=0,
		\\
		\ddot{\eta}+(3H-\frac{m^{2}c^{2}}{3}\mathcal{S}_{0})\dot{\eta}=0.
	\end{gather}
\end{subequations}
The above equation can be rewritten as
\begin{align}
    2\dot{H}+3H^{2}=\frac{8\pi G}{c^{2}}(p+P_{DE}),
    \\
    1=\frac{8\pi G}{3H^{2}}(\rho+\rho_{DE}),
\end{align}
and
\begin{subequations}
    \label{s3}
    \begin{align}
        P_{DE}&=\frac{m^{2}c^{4}}{24\pi G}(H\mathcal{S}_{0}-U),
        \\
        \rho_{DE}&=\frac{m^{2}c^{2}}{24\pi G}(U-\dot{\mathcal{S}}_{0}),
        \\
        w_{DE}&=\frac{P_{DE}}{\rho_{DE}c^{2}}.
    \end{align}
\end{subequations}
To solve this set of equations, we introduce 6 dimensionless variables
\begin{align}
    x_{1}&=\frac{m^{2}c^{2}U}{9H^{2}},\quad x_{2}=\frac{m^{2}c^{2}\dot{U}}{9H^{3}},\quad x_{3}=\frac{m^{2}c^{2}\mathcal{S}_{0}}{9H}, \notag
    \\
    x_{4}&=\frac{m^{2}c^{2}\dot{\mathcal{S}}_{0}}{9H^{2}},\quad x_{5}=\frac{\Sigma}{3H^{2}},\quad x_{6}=\frac{m^{2}c^{2}}{H^{2}},
\end{align}
along with the dimensionless parameter
\begin{equation}
	\gamma=\frac{2\dot{H}}{3H^{2}}=-(1-x_{1}+x_{3}+x_{5})-w(1-x_{1}+x_{4}-x_{5}).
	\label{gamma}
\end{equation}
With these variables, Eq. (\ref{s2}) becomes
\begin{subequations}
    \label{xn}
    \begin{align}
        \frac{\mathrm{d}x_{1}}{\mathrm{d}N}&=x_{2}-3\gamma x_{1},
        \\
        \frac{\mathrm{d}x_{2}}{\mathrm{d}N}&=(\gamma+\frac{4}{3})x_{6}-(3+\frac{9\gamma}{2})x_{2}+\frac{2}{3}x_{6}x_{5},
        \\
        \frac{\mathrm{d}x_{3}}{\mathrm{d}N}&=x_{4}-\frac{3}{2}\gamma x_{3},
        \\
        \frac{\mathrm{d}x_{4}}{\mathrm{d}N}&=x_{2}+3x_{3}-3(1+\gamma)x_{4}+6x_{3}x_{5},
        \\
        \frac{\mathrm{d}x_{5}}{\mathrm{d}N}&=3(2x_{3}-\gamma -2)x_{5},
        \\
        \frac{\mathrm{d}x_{6}}{\mathrm{d}N}&=-3\gamma x_{6},
    \end{align}
\end{subequations}
where $N=\ln\left(a/a_{0}\right)$ and $a_{0}$ is the scale factor at present. Equation (\ref{s3}) can be rewritten as 
\begin{subequations}
    \label{s4}
    \begin{align}
        P_{DE}&=\frac{3m^{2}c^{4}}{8\pi G}\frac{(x_{3}-x_{1}+x_{5})}{x_{6}}\label{sp},
        \\
        \rho_{DE}&=\frac{3m^{2}c^{2}}{8\pi G}\frac{(x_{1}-x_{4}+x_{5})}{x_{6}}\label{srho},
        \\
        w_{DE}&=\frac{x_{3}-x_{1}+x_{5}}{x_{1}-x_{4}+x_{5}}.
    \end{align}
\end{subequations}

\section{PHASE SPACE ANALYSIS}\label{sec:03}
To learn the overall properties and stability of the system, this section undertakes a phase-space analysis. Critical points will be identified, and trajectories around these points will be analyzed.  

Setting the right-hand side (RHS) of Eq.~(\ref{xn})  to zero yields the critical points. As a crucial component of phase-space analysis, the stability of critical points reflects the evolution and eventual state of a system. Critical points fall into three categories: stable, unstable, and saddle. The determination of a critical point's class is achieved by computing the eigenvalues of the Jacobian matrix. Table~\ref{b1} presents the critical points of the six-dimensional dynamic system along with the stability classification for each critical point. Notably, there exists a unique stable critical point, $P_{9}$. Nevertheless, it acts merely as a local attractor, so only a subset of initial conditions asymptotically approach it. As can be seen in the following discussion, Fig. \ref{t2} and Eq. (\ref{eq:3.8}), trajectories originating outside the region enclosed by the yellow curve diverge and reach infinity within a finite time.
\begin{table*}[t]
	\begin{center}
    	\caption[crit]{The critical points of the six-dimensional dynamical system (\ref{xn}). ``$+$'', ``$-$'', and ``$0$'' indicate positive, negative, and zero eigenvalues, respectively;
``$+\mathrm{(Re)}$'' denotes a complex eigenvalue with positive real part.}\label{b1}
    	\renewcommand{\arraystretch}{1.1}
    	\setlength{\tabcolsep}{8mm}
    	{
    		\begin{tabular}{c|c|c|c}
    			\hline 
    			\hline 
    			label & ($x_{1}$ , $x_{2}$ , $x_{3}$ , $x_{4}$ , $x_{5}$ , $x_{6}$) & eigenvalue & stability  \\ 
    			\hline
    			$P_{1}$ & (0, 0, $1/2$, $-3/4$, $-1/2$, 0) & $+ + +\mathrm{(Re)}$ $+\mathrm{(Re)} +\ -$ & saddle \\
    			$P_{2}$ & ($2/3$, $-4/3$, $2/3$, $-2/3$, $-1/3$, 0) & $- + + +\mathrm{(Re)}$ $+\mathrm{(Re)}\ -$ & saddle\\
    			$P_{3}$ & ($3/2$, 0, 1, 0, $-1/2$, 0) & $- - - +\mathrm{(Re)}$ $+\mathrm{(Re)}\ 0$ & saddle\\
    			$P_{4}$ & (0, 0, 0, 0, 1, 0) & $+ + + + +\ 0$ & saddle\\
    			$P_{5}$ & (1, 0, 0, 0, 0, 0) & $- - - - +\ 0$ & saddle\\ 
    			$P_{6}$ & (0, 0, 0, 0, 0, 0) & $- + + + +\ -$ & saddle \\
    			$P_{7}$ & ($2/3$, $-4/3$, $1/3$, $-1/3$, 0, 0) & $- + - + +\ -$ & saddle\\
    			$P_{8}$ & (0, 0, $(-3+\sqrt{33})/6$, $-1$, 0, 0) & $+ + + + +\ +$ & unstable node \\
    			$P_{9}$ & (0, 0, $(-3-\sqrt{33})/6$, $-1$, 0, 0) & $- - - - -\ -$ & stable node\\
    			\hline 
    			\hline
    		\end{tabular}
    	}
	\end{center}
\end{table*}

\subsection{THE 3D PHASE-SPACE ANALYSIS}\label{sec:03a}
Analyzing the full six-dimensional phase-space is inherently challenging, and it is not feasible to directly provide a distribution diagram for it. Fortunately, we find that there exists a series of closed subspace of this system --- initial conditions on the subspace will always remain on it. We defer the analysis of the full six-dimensional dynamical system to Sec.~\ref{sec:04}. Here, we conduct our analysis within the closed subspaces. Setting $x_1=x_2=x_6=0$ gives the three-dimensional subspace. The dynamical equations are simplified to be
\begin{subequations}
	\label{xd3}
	\begin{align}
		\frac{\mathrm{d}x_{3}}{\mathrm{d}N}&=x_{4}-\frac{3}{2}\gamma x_{3},
		\\
		\frac{\mathrm{d}x_{4}}{\mathrm{d}N}&=x_{2}+3x_{3}-3(1+\gamma)x_{4}+6x_{3}x_{5},
		\\
		\frac{\mathrm{d}x_{5}}{\mathrm{d}N}&=3(2x_{3}-\gamma -2)x_{5},
	\end{align}
\end{subequations}
where Eq. (\ref{gamma}) becomes
\begin{equation}
	\gamma=\frac{2\dot{H}}{3H^{2}}=-(1+x_{3}+x_{5})-w(1+x_{4}-x_{5}).
\end{equation}
Note that the $x_{5}$ coordinate reflects the anisotropy of the universe under discussion. We only consider pressureless fluid and thus $w=0$.

\begin{figure}[t]  
	\centering
	\includegraphics[width=1\linewidth]{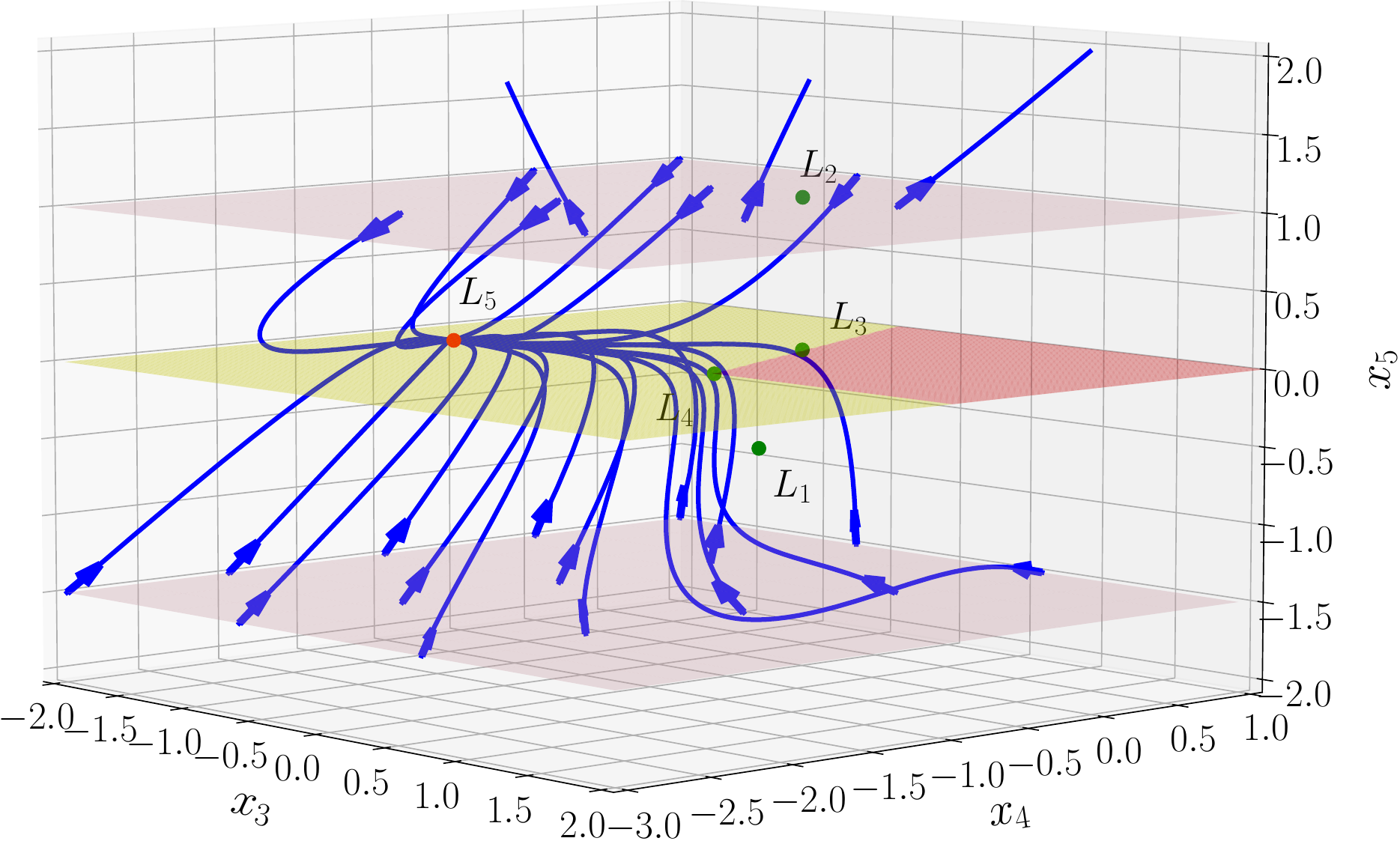}
	\caption{The trajectories in the three-dimensional phase space are shown, illustrating the evolutions governed by Eq. (\ref{xd3}) with initial conditions on the planes $x_{5} = 1$ and $x_{5} = -1.5$. Points $L_1$ to $L_5$ denote critical points, with green points indicating unstable points and red point indicating stable point. The significance of the red-shaded region is discussed in the main text.}  
	\label{t1}
\end{figure}

Through calculations, we arrive at Table \ref{b2}, which indicates the existence of a stable point $L_{5}$, which corresponds to $P_9$ in Table \ref{b1}.
Fig. \ref{t1} shows the three-dimensional phase portrait. It can be observed that, within this selected subspace, the trajectories corresponding to most initial conditions ultimately converge toward the plane $x_{5} = 0$, indicating the presence of a local attracting plane in this subspace. The yellow region on this plane will converge to the stable point $L_{5}$, while points in the red region will diverge away (see Sec.~\ref{sec:03b} for details).
\begin{table*}[t]
	\begin{center}
		\caption[crit]{The critical points of the three-dimensional dynamical system (\ref{xd3}). }\label{b2}
		\renewcommand{\arraystretch}{1.1}
		\setlength{\tabcolsep}{8mm}
		{
			\begin{tabular}{c|c|c|c}
				\hline 
				\hline 
				label & ($x_{3}$ , $x_{4}$ , $x_{5}$) & eigenvalue & stability  \\ 
				\hline
				$L_{1}$ & ($1/2$, $-3/4$, $-1/2$) & $+\mathrm{(Re)}$ $+\mathrm{(Re)} \ -$ & saddle \\
				$L_{2}$ & (0, 0, 1) & $+\  +\ 0\ $ & saddle\\ 
				$L_{3}$ & (0, 0, 0) & $-\  +\ -$ & saddle \\
				$L_{4}$ & ($(-3+\sqrt{33})/6$, $-1$, 0) & $+\  +\  +$ & unstable \\
				$L_{5}$ & ($(-3-\sqrt{33})/6$, $-1$, 0) & $-\  -\  -$ & stable\\
				\hline 
				\hline
			\end{tabular}
		}
	\end{center}
\end{table*}

\subsection{THE 2D PHASE-SPACE ANALYSIS}\label{sec:03b}
\begin{figure}[t]
 	\centering
 	\includegraphics[width=1\linewidth]{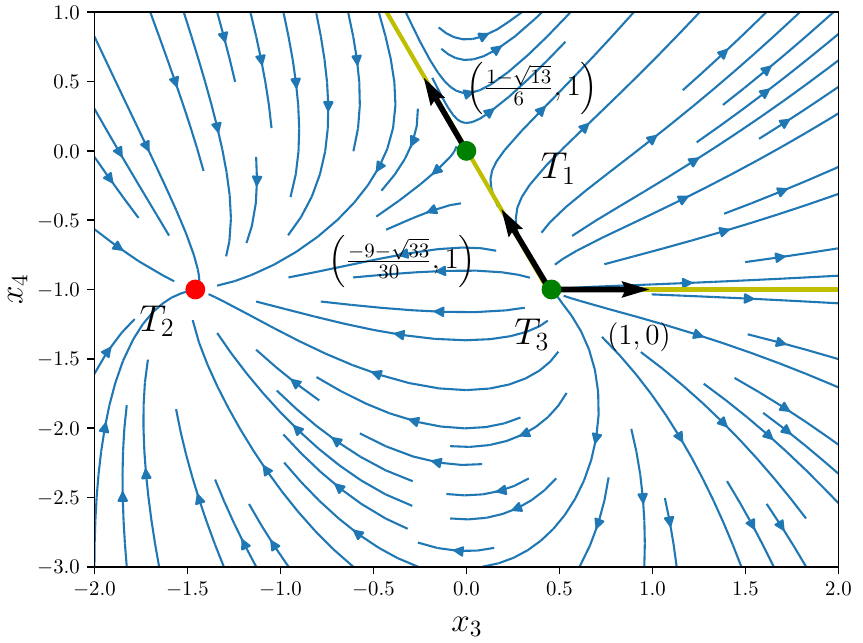}
 	\caption{Phase portrait of the dynamical system (\ref{2w}). The critical points of the dynamical system are marked by red and green symbols, where the red point is stable and the two green points are unstable. The yellow curve indicates the asymptote: trajectories starting below it converge to the stable point (red), whereas those starting above it diverge. The black arrows show the eigenvector direction at the unstable point, which is tangent to the yellow curve at that location.} 
 	\label{t2}
 \end{figure}

From Eq. (\ref{xn}) , we know that points on the $x_{3}-x_{4}$ plane will always remain on the $x_{3}-x_{4}$ plane, which plane also corresponds to the local converging plane depicted in Fig. \ref{t1}. Therefore, we can focus on this cross-section of the six-dimensional system and present the two-dimensional phase-space distribution for that cross-section. Noting that the stable point $P_{9}$ lies within this special plane, the above discussion remains general and effectively reflects the evolution of the 6-dimensional system around the stable point $P_{9}$. The two-dimensional dynamical equations read
\begin{subequations}
	\begin{align}
    	\frac{\mathrm{d}x_{3}}{\mathrm{d}N}&=x_{4}-\frac{3}{2}\gamma x_{3},
    	\\
    	\frac{\mathrm{d}x_{4}}{\mathrm{d}N}&=3x_{3}-3(1+\gamma)x_{4},
    \end{align}
	\label{2w}
\end{subequations}
where 
\begin{equation}
      \gamma=-1-x_{3}-w(1+x_{4}).
\end{equation}
Taking $w=0$, the same analysis leads to Table \ref{b3}. Figure \ref{t2} shows the phase-space evolution of $x_{3}$ and $x_{4}$. First, $T_{2}$ is a stable node in this subspace. In addition, a discernible dividing line traverses through the point $T_{3}$, marked with a yellow line in Fig. \ref{t2}. In this subspace, trajectories starting below this line converge to $T_{2}$, whereas those starting above it diverge (the red and yellow regions on the $x_{5} = 0$ plane in Fig. \ref{t1} correspond to this). By numerically estimating the slope of trajectories in the vicinity of the dividing line and comparing it with the eigenvector direction, we find that the eigenvector at points on the dividing line is tangent to the dividing line.  

 \begin{table*}[t]
 	\begin{center}
 		\caption[crit]{The critical points of the two-dimensional dynamical system (\ref{2w}). ``$+$'' and ``$-$'' denote the sign of eigenvalues.}\label{b3}
 		\renewcommand{\arraystretch}{1.1}
 		\setlength{\tabcolsep}{8mm}
 		{
 			\begin{tabular}{c|c|c|c}
 				\hline 
 				\hline
 				label & ($x_{3}$ , $x_{4}$) & eigenvalue & stability  \\ 
 				\hline 
 				$T_{1}$ & (0 , 0 ) & $+\ -$  & saddle \\
 				$T_{2}$ & ($(-3-\sqrt{33})/6$  , -1 ) & $-\ -$  & stable\\
 				$T_{3}$ & ($(-3+\sqrt{33})/6$ , -1) & $+\ +$  & unstable \\
 				\hline 
 				\hline
 			\end{tabular}
 		}
 	\end{center}
 \end{table*}
 
Further setting $x_4=-1$ gives a one-dimensional closed subspace, and the dynamical equation is
\begin{equation}
    \frac{\mathrm{d}x_{3}}{\mathrm{d}N}=\frac{3}{2}x_{3}^{2}+\frac{3}{2}x_{3}-1
    \label{x3},
\end{equation}
which reflects the dynamics of the horizontal axis. This is the Riccati's equation. An analytical solution that describe the evolution from $T_3$ to $T_2$ is
\begin{equation}
    x_{3}=-\frac{\sqrt{33}\left[\sqrt{33}+11\mathrm{tanh}\left(\frac{\sqrt{33}}{4}a+\frac{33}{4}N\right)\right]}{66}.
\end{equation}
where $a$ is an integration constant determined by the initial conditions. To examine the evolution on the right-hand side of the critical point $T_{3}$, we further simplify Eq. (\ref{x3}). As $x_{3}$ increases, the right-hand side of Eq. (\ref{x3}) becomes dominated by the highest-order term. For sufficiently large $x_{3}$, we can rewrite it as
\begin{equation}
    \frac{\mathrm{d}x_{3}}{\mathrm{d}N}=\frac{3}{2}x_{3}^{2}.
\end{equation}
The above equation can be solved analytically as,
\begin{equation}\label{eq:3.8}
    x_{3}=-\frac{2}{3N+2b}.
\end{equation}
and $b$ an the integration constant. The solution becomes singular at $N=-2b/3$: as $N$ approaches $-2b/3$, $x_{3}$ diverges to infinity. This implies that trajectories on the right-hand side of $T_{3}$ with $x_{4}=-1$ will diverge in finite time.

\section{NUMERICAL EVOLUTION}\label{sec:04}
\begin{figure}[t]  
	\centering
	\includegraphics[width=1\linewidth]{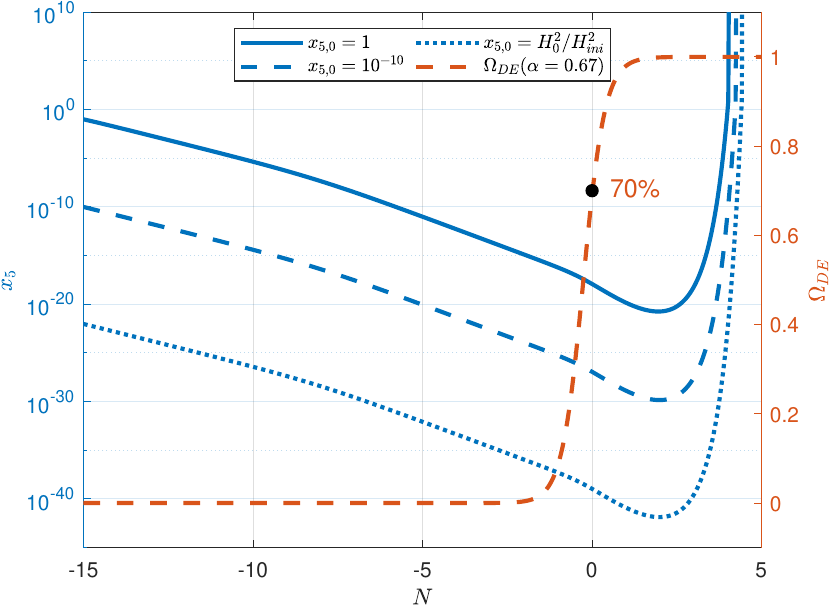}
	\caption{Orange part: The orange dashed curve shows the evolution of $\Omega_{DE}$ as a function of $N$. Here, $N=0$ corresponds to the present epoch. The black point marks the present value of $\Omega_{DE}$ when $\alpha=0.67$, consistent with current observations (Sec.~\ref{sec:03}).
		Blue part: The blue curves show the evolution of $x_{5}$ for different values of $x_{5,0}$. The initial conditions are given by $x_{1}=0$, $x_{2}=0$, $x_{3}=0$, $x_{4}=0$, $x_{5}=x_{5,0}$, and $x_{6}=\alpha^{2} H_{0}^{2}/H^{2}$, where $x_{5,0}$ sets the initial values of $x_{5}$. Since $x_{5}$ diverges during the evolution, the plot is truncated for clarity.}  
	\label{t3}
\end{figure}

In this section, we will give the numerical evolution of the dynamical system. It is worth emphasizing that, in contrast to the subspaces considered previously, we adopt here a different subspace that is more directly motivated by observational considerations. We first consider a special case of $x_{5}= 0$, which corresponds to the isotropic universe. The total equation of state for radiation and matter can be written as
\begin{equation}
	w=\frac{1}{3}\left(1+\mathrm{e}^{N-N_{eq}}\right)^{-1}.
\end{equation}
We choose $H_{0}=70$ km/s/Mpc and $N_{eq}=-8.13$ \citep{2020AA641A6P}. Tuning the free mass parameter $m$ (or, equivalently, tuning $\alpha$, as $m=\alpha H_0/c$), we obtain the orange part of Fig.~\ref{t3}.
It is evident that when $m=0.67H_{0}$, $\Omega_{DE}=0.7$ aligns well with observational data.

Nonzero $x_5$ corresponds to an anisotropic Universe. As mentioned in Sec. \ref{sec:02}, both $\Sigma$ and $x_{5}$ serve as indicators of the cosmic anisotropy. However, a pertinent question arises: which variable more accurately depicts the evolution of this anisotropy? By virtue of Eqs. (\ref{sig1}) - (\ref{sig3}), it becomes apparent that $\Sigma$ is associated with the expansion rate of the universe, whereas $x_{5}$ is normalized by the square of the expansion rate. Hence, for the ensuing discussion, we shall select $x_{5}$ as the preferred variable to investigate the evolution of cosmic anisotropy. Solving Eq. (\ref{xn}) numerically, we obtain the blue part of Fig. \ref{t3}. To examine the impact of the initial value of $x_{5}$, we repeat the calculation for various initial conditions of $x_{5}$. Considering that $x_{5}$ is inversely proportional to $H^{2}$ and is only one coefficient different from $x_{6}$, we also try to take the initial value of $x_{5}$ as $H_{0}^{2}/H^{2}$. 
From the figure, it is evident that the anisotropy is currently in a decreasing stage, and inflection points are expected to appear in the future, after which the anisotropy gradually diverge with time. Different initial values of $x_{5}$ do not significantly impact the overall evolutionary trend, while the occurrence time of the inflection point is slightly delayed as the initial value decreases. The divergent behavior observed in Fig. \ref{t3} is consistent with Eq. (\ref{eq:3.8}).

\begin{figure}[t] 
	\centering
	\includegraphics[width=0.9\linewidth]{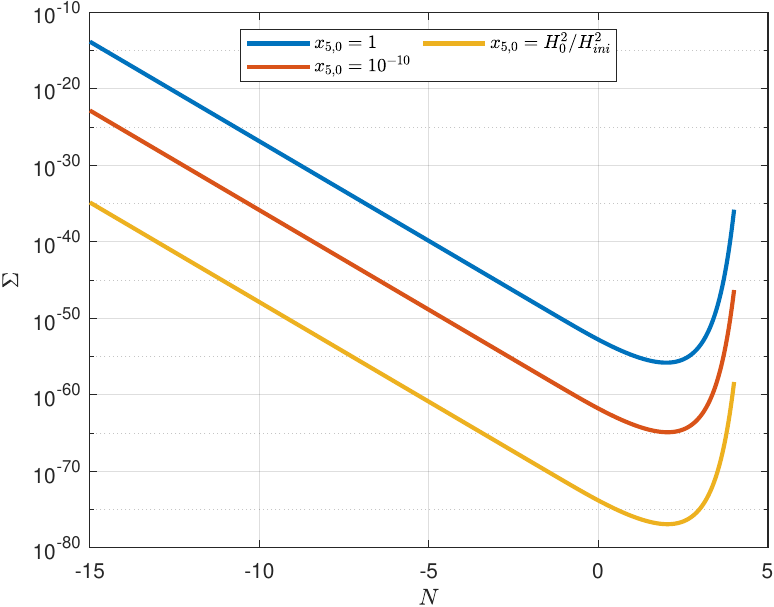}
	\caption{Evolution of $\Sigma$ with different $x_{5,0}$. The initial conditions are given by $x_{1}=0$, $x_{2}=0$, $x_{3}=0$, $x_{4}=0$, $x_{5}=x_{5,0}$, and $x_{6}=\alpha^{2} H_{0}^{2}/H^{2}$, where $x_{5,0}$ sets the initial values of $x_{5}$.} 
	\label{t4}
\end{figure}

As a complement to the previous discussion, we further examine the evolution of the anisotropy variable $\Sigma$, which can be expressed as
\begin{equation}
    \Sigma=\frac{3m^{2}c^{2}x_{5}}{x_{6}}=\kappa \frac{x_{5}}{x_{6}}\label{sigma},
\end{equation}
where $\kappa=3m^2c^2=3\alpha^2H_0^2$. Similar to the analysis of $x_{5}$, we employ the same set of initial conditions to obtain the numerical solutions. Without loss of generality, we set $\alpha=0.67$. By solving Eq. (\ref{xn}) numerically and substituting the results into Eq. (\ref{sigma}), we obtain Fig. \ref{t4}. It is evident that $\Sigma$ likewise grows and eventually diverges at late times, mirroring the evolutionary trend previously noted for $x_{5}$.

\section{CONCLUSION}\label{sec:06}
The cosmological equations describing cosmic dynamics typically form systems of ordinary differential equations, and one of the most elegant ways to investigate them is to cast them into the form of dynamical systems. This allows the use of powerful analytical and numerical methods to gain a quantitative understanding of the cosmological dynamics derived by the models under study \citep{BAHAMONDE20181}. In this paper, we provide a dynamical analysis of nonlocal RT gravity in a Bianchi type-I spacetime. By introducing six dimensionless variables, we derive the dynamical system (Eq.~(\ref{xn})) from the cosmological equations. The phase-space analysis reveals the existence of a unique stable critical point, denoted as $P_{9}$ (see Table \ref{b1}). The numerical results, particularly the late-time growth of the anisotropy variable $x_{5}$, demonstrate that unlike most dark energy models, where anisotropy typically decays with time, Bianchi type-I anisotropy is enhanced once the Universe enters an accelerating phase in nonlocal RT gravity. This indicates a violation of the cosmic no-hair theorem in the nonlocal RT gravity: a de Sitter-like accelerated regime does not necessarily erase anisotropies. Since the FLRW metric is recovered as the isotropic limit of the Bianchi type I spacetime, the growth of anisotropy indicates a Bianchi instability of the FLRW metric within nonlocal RT gravity.

\section*{Acknowledgements}
This work was supported by the National Natural Science Foundation of China under Grant Nos.~12405050 and 12433001, and the Fundamental Research Funds for the Central Universities.

\bibliographystyle{elsarticle-num} 

\end{document}